\documentclass[twocolumn]{aastex631}%linenumbers

\usepackage[utf8]{inputenc}
\usepackage[T1]{fontenc}
\usepackage{soul,hyperref}
\usepackage{slantsc}
\usepackage{upgreek}

\newcommand{\mhi}{M_\mathrm{HI}}
\newcommand{\hi}{H{\sc i}}

\newcommand{\unit}[1]{\,\mathrm{#1}}

\newcommand{\milli}{\ensuremath{\mathrm{m}}}
\newcommand{\giga}{\ensuremath{\mathrm{G}}}
\newcommand{\micro}{\ensuremath{\mu}}

\newcommand{\arcsecond}{\ensuremath{^{\prime\prime}}}
\newcommand{\Hz}{\ensuremath{\mathrm{Hz}}}
\newcommand{\watt}{\ensuremath{\mathrm{W}}}
\newcommand{\kms}{\ensuremath{\mathrm{km\,s^{-1}}}}
\newcommand{\cmtwo}{\ensuremath{\mathrm{cm^{-2}}}}
\newcommand{\meter}{\ensuremath{\mathrm{m}}}
\newcommand{\pc}{\ensuremath{\mathrm{pc}}}
\newcommand{\kpc}{\ensuremath{\mathrm{kpc}}}
\newcommand{\mpc}{\ensuremath{\mathrm{Mpc}}}
\newcommand{\msun}{\ensuremath{\mathrm{M_\odot}}}
\newcommand{\yr}{\ensuremath{\mathrm{yr}}}
\newcommand{\beam}{\ensuremath{\mathrm{beam}}}
\newcommand{\jansky}{\ensuremath{\mathrm{Jy}}}
\newcommand{\ujyb}{\ensuremath{\mathrm{\mu Jy\,beam^{-1}}}}
\newcommand{\mjyb}{\ensuremath{\mathrm{mJy\,beam^{-1}}}}

\newcommand{\SI}[2]{#1\,#2}
\newcommand{\si}[1]{\unit{#1}}
\newcommand{\SIrange}[3]{#1--#2\,#3}

\newcommand{\per}[1]{#1^{-1}}
\newcommand{\persquare}[1]{#1^{-2}}
\newcommand{\e}[1]{\times10^{#1}}
\usepackage{booktabs}
\begin{document}

\title{WALLABY pilot survey: Blinded by the light -- discovery of a fourth member in the ESO 179-013 system}

\author[0000-0003-0634-7710]{R. Guimar\~aes Silva}
\correspondingauthor{R. Guimar\~aes Silva}
\email{rgs@astro.ufrj.br}
\affiliation{Observat\'orio do Valongo, Universidade Federal do Rio de Janeiro, Ladeira Pedro Antônio 43, 20080-090 Rio de Janeiro, RJ, Brazil}

\author[0000-0003-4675-3246]{M. Grossi}
\affiliation{Observat\'orio do Valongo, Universidade Federal do Rio de Janeiro, Ladeira Pedro Antônio 43, 20080-090 Rio de Janeiro, RJ, Brazil}

\author[0000-0001-9388-7146]{D. R. Gonçalves}
\affiliation{Observat\'orio do Valongo, Universidade Federal do Rio de Janeiro, Ladeira Pedro Antônio 43, 20080-090 Rio de Janeiro, RJ, Brazil}

\author[0000-0002-5788-2628]{E. Corbelli}
\affiliation{INAF - Osservatorio Astrofisico di Arcetri, L. E. Fermi 5, 50125, Firenze, IT}

\author{B. Catinella}
\affiliation{ICRAR, The University of Western Australia, 35 Stirling Highway, Crawley WA 6009, Australia}

\author{N. Deg}
\affiliation{Department of Physics, Engineering Physics, and Astronomy, Queen's University, Kingston, ON, Canada, K7L 3N6}

\author{B. W. Holwerda}
\affiliation{University of Louisville, Department of Physics and Astronomy, 102 Natural Science Building, 40292 KY Louisville, USA}

\author{R. Ianjamasimanana}
\affiliation{Instituto de Astrofísica de Andalucía (CSIC), Glorieta de la Astronomía s/n, 18008 Granada, Spain}

\author{D. A. Leahy}
\affiliation{Department of Physics and Astronomy, University of Calgary, 2500 University Dr. NW, Calgary, AB, T2N 1N4, Canada}

\author{P. E. Mancera Piña}
\affiliation{Leiden Observatory, Leiden University, PO Box 9513, 2300 RA, Leiden, the Netherlands}

\author[0000-0002-7607-081X]{S. Sankar}
\affiliation{ICRAR, The University of Western Australia, 35 Stirling Highway, Crawley WA 6009, Australia}

\author{K. Spekkens}
\affiliation{Department of Physics, Engineering Physics, and Astronomy, Queen's University, Kingston, ON, Canada, K7L 3N6}

\author[0000-0001-9414-175X]{S. F. ur Rahman}
\affiliation{Syed Babar Ali School of Science and Engineering, Lahore University of Management Sciences, Lahore, Pakistan}
%\affiliation{Lahore University of Management Sciences (LUMS), Lahore, Pakistan, 44} 
%\affiliation{NCBC at NED University of Engineering and Technology, Karachi, Pakistan}

\author{T. Westmeier}
\affiliation{ICRAR, The University of Western Australia, 35 Stirling Highway, Crawley WA 6009, Australia}

\author{O. I. Wong}
\affiliation{Australia Telescope National Facility, CSIRO, Space and Astronomy, PO Box 1130, Bentley WA 6102, Australia
}
\affiliation{ICRAR, The University of Western Australia, 35 Stirling Highway, Crawley WA 6009, Australia}



\begin{abstract}
We present new ASKAP/WALLABY \hi\ observations of the nearby dwarf galaxy system ESO 179-013 (Kathryn's Wheel), the nearest known collisional ring galaxy, located $\sim\SI{10}{\mpc}$ away in the Local Void. The system is composed of three previously known dwarf galaxies embedded in a large \hi\ envelope, with a newly discovered fourth member identified through \hi\ and radio continuum emission behind a bright foreground binary. Galaxy D exhibits the highest star formation rate in the group and deviates from the \hi\ mass-diameter relation, suggesting it is a compact, gas-rich dwarf missed due to stellar foreground contamination. The \hi\ data reveal for the first time an extended \hi\ envelope around the whole system, the neutral gas counterpart of the star-forming ring and gas bridges among members, suggesting a more complex interaction history than the previously proposed collisional ring scenario. ESO 179-013 thus provides a rare opportunity to study hierarchical assembly and gas dynamics in underdense environments and demonstrates the power of blind \hi\ surveys in identifying faint members of low-mass compact groups.
\end{abstract}



%%%%%%%%%%%%%%%%%%%%%%%%%%%%%%%%%%%%%%%%%%%%%%%%%%

%%%%%%%%%%%%%%%%% BODY OF PAPER %%%%%%%%%%%%%%%%%%

\section{Introduction}\label{sec:intro}

Collisional ring galaxies (CRGs) are formed after galaxy-galaxy collisions when a compact ``bullet'' hits a ``target'' disk galaxy in a drop-through collision. This collision generates a radially propagating density wave that displaces gas and stars, triggering star formation along rings \citep{lyndsInterpretationRingGalaxies1976}. CRGs represent an extreme mode of interaction that is unlike mergers but can be used to understand the nature of the collision itself. The shape and strength of the gravitational potential of the target galaxy can be probed through the resulting collisional ring \citep[e.g,][]{fiacconiAdaptiveMeshRefinement2012}. Moreover, due to the symmetry or ``simplicity'' of the collision, it is possible to predict how these systems evolve over time. They have been proposed as the progenitors of Giant Low Surface Brightness Galaxies (GLSBs), which are not only rare but also difficult to explain with the standard $\Lambda$CDM cosmology \citep{mapelliAreRingGalaxies2008}. One of the main difficulties in studying these types of systems is that the duration of the ring is very brief -- a few hundred million years after the collision \citep[e.g.,][]{renaudMorphologyEnhancedStar2018}. 

Within this framework, the triple system ESO 179-013, known as Kathryn's Wheel \citep{parkerKathrynsWheelSpectacular2015} contains the closest known collisional ring at a distance of $\sim\SI{10}{\mpc}$ and is composed of three known low-mass galaxies ($M_{*} < \SI{3\e9}{\msun}$). Beyond the rare phenomenon of a collisional ring, this system also offers the opportunity to study the interaction between at least three dwarf galaxies in an environment where cosmological simulations predict that dwarf triplets are rare \citep{beslaFrequencyDwarfGalaxy2018}. 

In \cite{parkerKathrynsWheelSpectacular2015}, ESO 179-013 had no identified nearby galaxies within 1 Mpc, and more recent works place it within the Local Void \citep{tullyCosmicflows3CosmographyLocal2019,courtoisWALLABYPrepilotPilot2023}. This is consistent with expectations, as CRGs are not expected to be found in dense environments such as clusters \citep{madoreAtlasCatalogCollisional2009}. Galaxies in voids are expected to be isolated from complex environmental processes that modify galaxies in higher-density environments, i.e., driven by secular processes rather than environmental ones. This is particularly relevant when studying the neutral gas, which is highly sensitive to external perturbations \citep[e.g,][]{boselliEnvironmentalEffectsLateType2006}.

This system remains relatively unexplored since the discovery of the star-forming ring in 2015 due to its location near the Galactic plane ($b = -8^{\circ}$) in a region contaminated by foreground stars and close to a bright visual binary pair. Recently, a few works using radio data have reported its observation \citep[e.g.,][]{murugeshanHiRingGalaxies2023} and a possible association with a $\gamma$-ray source \citep{paliyaGRayemittingCollisionalRing2024}.

In this paper, we report new results from observations of the neutral atomic hydrogen (\hi) in this system using the Australian Square Kilometre Array Pathfinder \citep[ASKAP;][]{hotanAustralianSquareKilometre2021}. This paper is organized as follows. In Section 2, we describe the ASKAP/WALLABY observations and the complementary optical data used in our analysis. Section 3 presents the main results, which we discuss in Section 4. We summarize our conclusions in Section 5.

\section{Data}
\subsection{ASKAP Observations}
\begin{figure*}[!ht]
	\begin{center}
		\includegraphics[width=\linewidth]{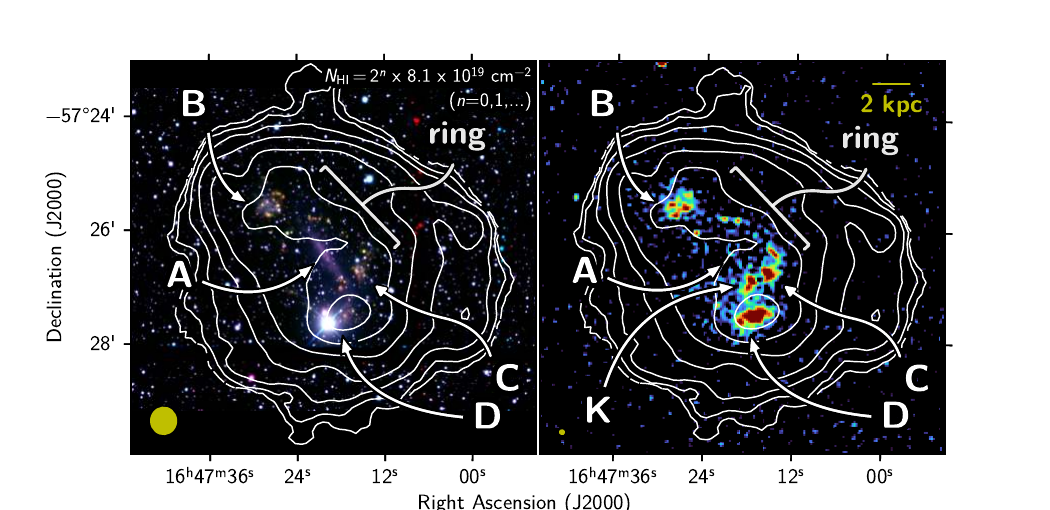}
	\end{center}
		\caption{\hi\ column density contours from ASKAP/WALLABY overlaid on a composite RGB image from CTIO/DECam on the left and on the 1.4 GHz radio continuum from ASKAP/WALLABY on the right. In the left image, red corresponds to H$\alpha$, green to [O\,\textsc{iii}], and blue to broadband $V$ (as in \citealt{parkerKathrynsWheelSpectacular2015}). The \hi\ contours are at column densities of $2^n \times 8.1 \times 10^{19}$ cm$^{-2}$ for $n = 0, 1, 2, \dots$, tracing the large-scale neutral gas envelope. The ASKAP synthesized beam is shown as a yellow filled circle in both panels: 30\arcsec\ for the \hi\ data and 6.8\arcsec$\times$6.3\arcsec\ for the radio continuum. The arrows indicate the positions of the galaxies (A, B, C, and D) and the ring.}\label{fig:radio}
	
\end{figure*}
ESO 179-013 was observed at 1.4 GHz with ASKAP  as part of the Wide-field ASKAP L-band Legacy All-sky Blind Survey \citep[WALLABY;][]{koribalskiWALLABYSKAPathfinder2020} pilot survey.  In the phase 1 of the pilot survey \citep{westmeierWALLABYPilotSurvey2022}, WALLABY observed three 60 deg$^2$ regions in the direction of the Hydra and Norma galaxy clusters and the NGC 4636 galaxy group with a nominal target root mean square (RMS) level of 1.6 mJy per 30$^{\prime\prime}$ beam and 18.5 kHz channel width. In these fields, the sources were found with the Source Finding Application \citep[SoFiA 2; ][]{westmeierSOFIA2Automated2021} and the main products of Data Release 1 (DR1) are a catalog of \hi\ sources and the corresponding data cube for each source \citep{westmeierWALLABYPilotSurvey2022}. For the analysis of ESO 179-013 (WALLABY J164720-572629) data we reran SoFiA 2 with the default parameters, except for the spatial kernels (we use 0, 3, 6, 9, 12, 15) and impose a flux threshold of 4$\times$RMS. The spatial kernels range is the same as \citet{obeirneWALLABYPilotSurvey2024}, albeit in steps of 3 pixels instead of 5 (1 pixel $\sim$ 0.3 kpc). We find that this scale of spatial smoothing combined with a higher flux threshold (WALLABY uses $3.5\times\mathrm{RMS}$ for source identification) ensures the recovery of extended gas emission while minimizing spurious detections for this particular system. The RMS in the data cube is \SI{1.8}{\mjyb}, corresponding to a 3$\sigma$ column density sensitivity of \SI{2.6e19}{\cmtwo} over \SI{3.9}{\kms}.

WALLABY data products also provide high-resolution radio continuum images at $\sim$ 1.4GHz \citep{koribalskiWALLABYSKAPathfinder2020}. %\citep{koribalskiWALLABYSKAPathfinder2020}.
For the ESO 179-013 field, the angular resolution of the radio continuum data is $\SI{6.8}{\arcsec}\times\SI{6.3}{\arcsec}$ with \SI{50}{\ujyb} RMS. Using these data, we aim to determine the star formation rate of the system components.

For the distance estimate, we adopt the systemic velocity reported by \cite{parkerKathrynsWheelSpectacular2015}, \SI{842}{\kms}, and compute the distance using the Cosmicflows-3 distance estimator \citep{kourkchiCosmicflows3TwoDistance2020},based on the CosmicFlows-3 catalog \citep{tullyCosmicflows32016} and available through the Extragalactic Distance Database\footnote{\url{http://edd.ifa.hawaii.edu/NAMcalculator/}}. The resulting flow-corrected distance to the system is $D = \SI{9.78}{\mpc}$. In this work, we adopt the following cosmological parameters: $H_0 = \SI{70}{\kms\per\mpc}$, $\Omega_m = 0.3$, and $\Omega_{\Lambda} = 0.7$.

\begin{figure*}
	\begin{center}
		\includegraphics[width=\linewidth]{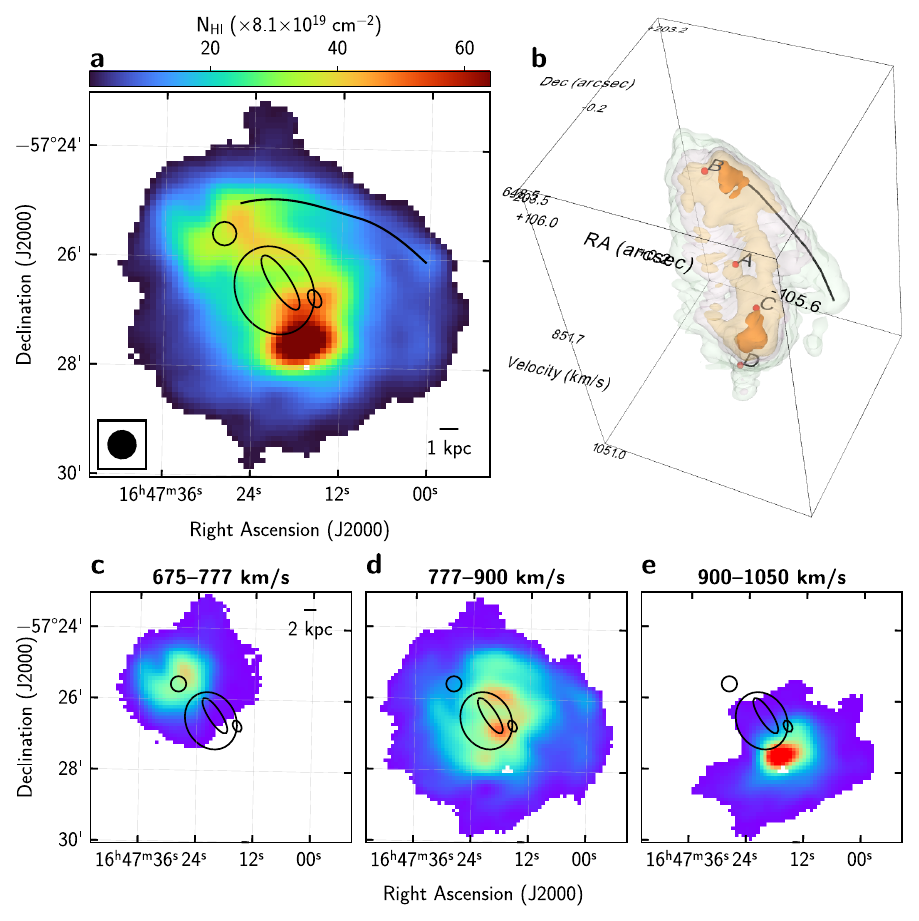}
	\end{center}
	\caption{\hi\ integrated  intensity (moment 0 map) of the ESO 179-013 system shown over different velocity ranges. The middle ellipse highlights galaxy A, the largest ellipse marks the ring, the northeastern circle identifies the galaxy B, and the small ellipse to the west indicates galaxy C.
	\textbf{(a)} Full velocity range, \SIrange{675}{1050}{\kms}, highlighting the extended gas envelope; 
	\textbf{(b)} three-dimensional visualization of ESO 179-013 with iso-surface levels of different column density of neutral gas, an interactive version of this figure is available in the online journal; 
	\textbf{(c)} \SIrange{675}{777}{\kms}, gas component of galaxy B; 
	\textbf{(d)} \SIrange{777}{900}{\kms}, emission from the ring, galaxy C, and the interaction among group members; 
	\textbf{(e)} \SIrange{900}{1050}{\kms}, emission dominated by galaxy D.
	Black ellipses mark the optical radii of the three known galaxies and the best fitting ellipse to the ring. The black continuous line indicates the plume possibly associated with galaxy B. The black filled circle in panel (a) is the beam.}\label{fig:summary}
\end{figure*}

\begin{figure*}
	\begin{center}
		\includegraphics[width=\linewidth]{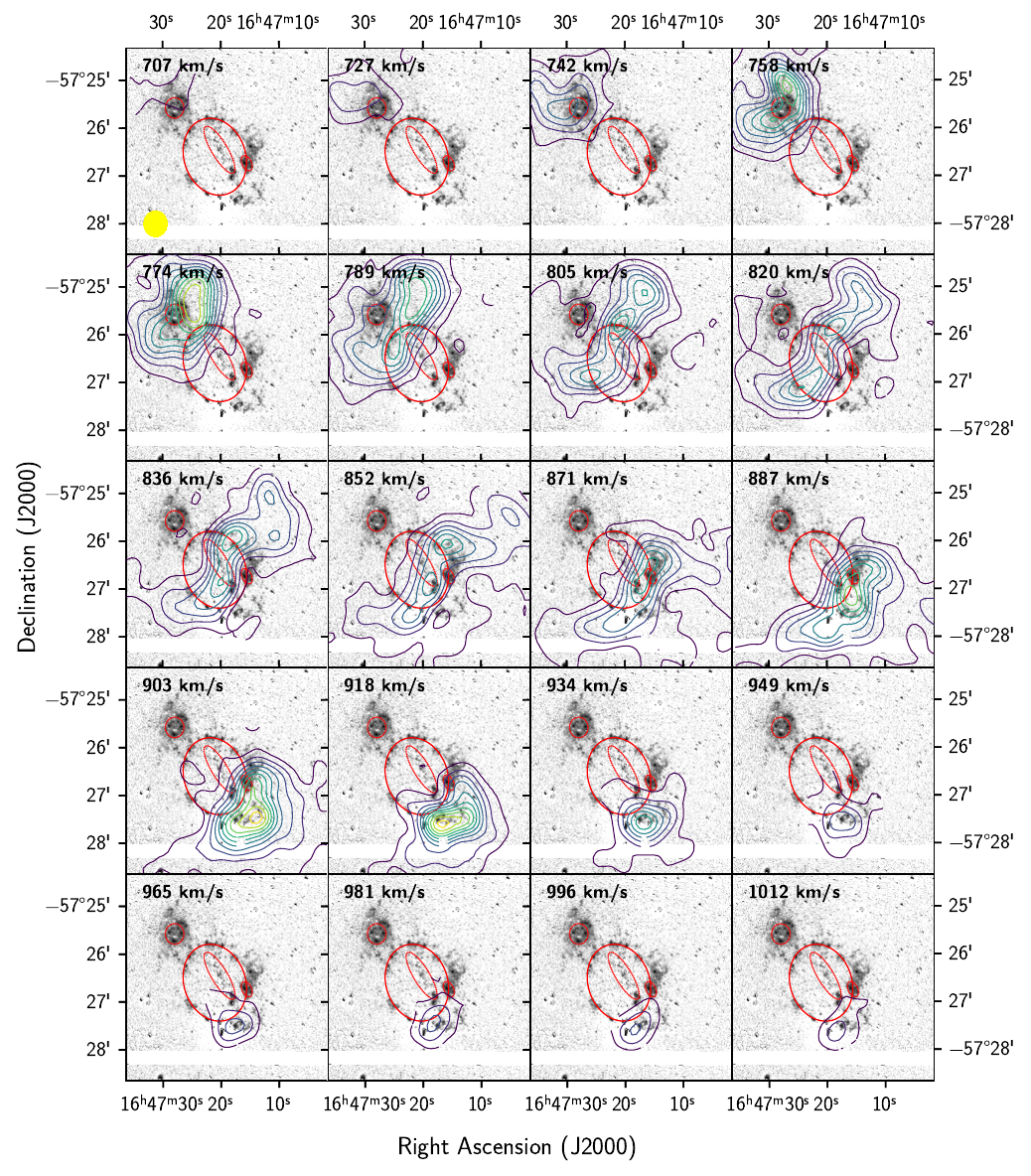}
	\end{center}
		\caption{Channel map of the \hi\ emission for ESO 179-013. Each panel displays a 415\arcsec$\times$415\arcsec region with contours of the \hi\ emission at different velocities. Contours are drawn at the $[2.6, 6.1,...,33.9]\times10^{19}\,\si{\cmtwo}$ levels. The red ellipses mark the optical positions of the ring and galaxies A, B, and C. The beam is shown as a yellow circle in the first panel. The background image is the continuum subtracted H$\alpha$ from CTIO \citep{parkerKathrynsWheelSpectacular2015} and the white strip marks a detector gap.}\label{fig:chmap}
\end{figure*}

\begin{figure*}
	\begin{center}
		\includegraphics[width=\linewidth]{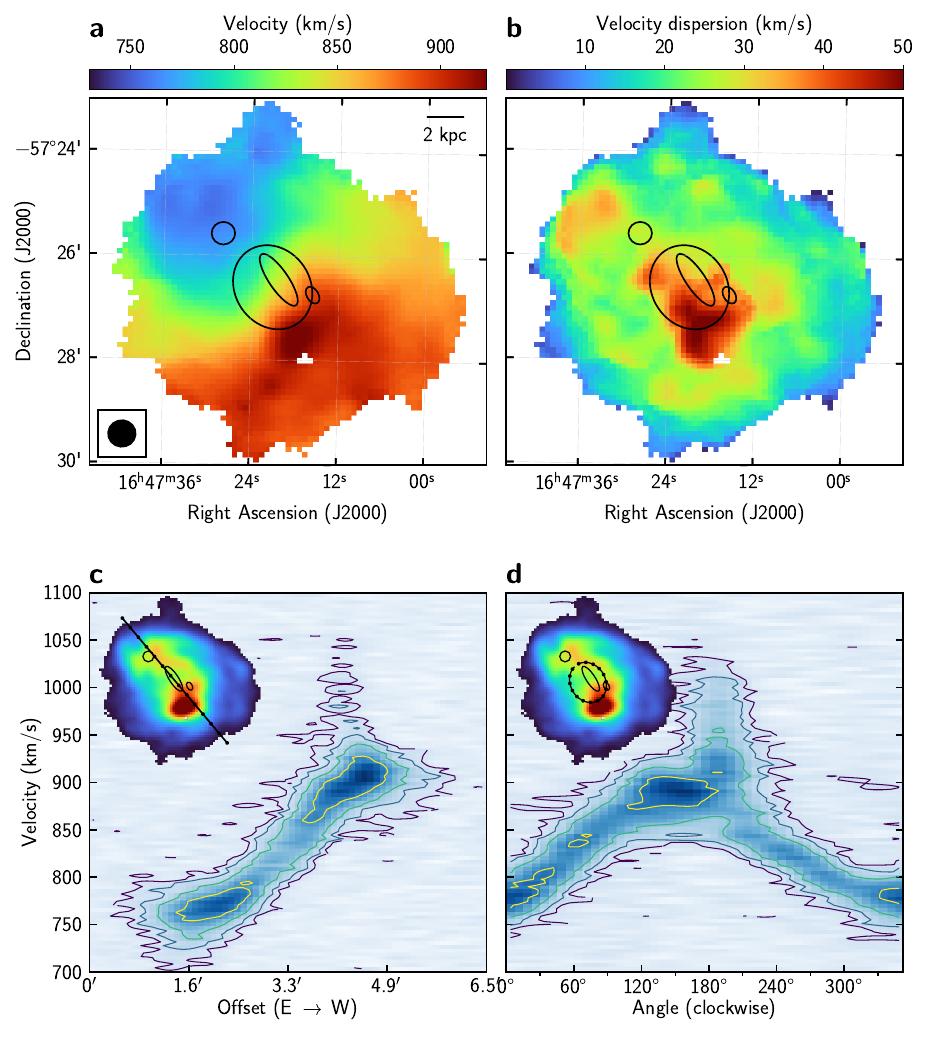}
	\end{center}	
		\caption{\hi\ kinematics of the ESO 179-013 system from ASKAP/WALLABY data;	\textbf{(a)} is the flux-weighted velocity field, and \textbf{(b)} is the velocity dispersion map. PV diagrams along the kinematic axis of the system \textbf{(c)}, and along the optical position of the ring \textbf{(d)}.  Galaxy D is detected as a vertical tail in both panels. The ring shows a sinusoidal pattern that suggests that rotation is the main component of the line-of-sight velocity. The largest ellipse marks the best-fitting optical ellipse of the H$\alpha$ ring, while the smaller ellipses indicate the optical extents of galaxies B, A, and C (from left to right). North is up and East is left. The black filled circle is the beam. The contour levels in the PV diagrams are $2^n \times 3\times\SI{1.8}{\mjyb}$ ($n =$ 0, 1, 2, 3).}
		\label{fig:vels}
\end{figure*}

%%% MOM 0

\section{Results}

\subsection{H\textsc{i} content and distribution}

In Figure \ref{fig:radio}, we show the \hi\ distribution obtained with SoFiA 2 from the WALLABY observations for ESO 179-013 overlaid on the optical (left) and radio continuum (right) images of the system. We highlight the position of the three known galaxies with arrows: WKK 7460 (hereafter galaxy A) is believed to be an edge-on dwarf with stellar mass comparable to the Large Magellanic Cloud; WKK 7463, (galaxy B) is the bullet and WKK 7457, (galaxy C) is the third member of the system, both with stellar masses comparable to the Small Magellanic Cloud \citep{parkerKathrynsWheelSpectacular2015}. The ring is visible as the bright H$\alpha$ knots distributed in a circle around galaxy A.

We show the \hi\ distribution in Figure \ref{fig:summary}a--e with overlays of the positions of the system components. We detect a large envelope of neutral hydrogen surrounding all the components of the system, extending out to $\sim\SI{9}{\kpc}$ from galaxy A down to a column density of $N_\mathrm{HI} = \SI{8e19}{\cmtwo}$. The total \hi\ mass of the system is $\mhi=\SI{2.25\pm 0.02\e9}{\msun}$ extending over a velocity range of $\sim\SI{400}{\kms}$. 
This value agrees with the earlier estimate of $\mhi = \SI{2\e9}{\msun}$ from \citet{koribalski1000BrightestHIPASS2004a} using HIPASS data. However, while \cite{parkerKathrynsWheelSpectacular2015} suggested that this \hi\ mass is likely associated with galaxy A, our data indicates otherwise.
Notably, the peak of the \hi\ distribution does not coincide with any of the three known galaxies of the system, but is instead located behind the bright foreground binary star (see Figure \ref{fig:radio}). This spatial offset, combined with the fact that the emission appears at a significantly higher systemic velocity than the rest of the system (Figure \ref{fig:summary}e), suggests the presence of an additional source (WALLABY J164717-572735, hereafter galaxy D), also visible as a bright source in the continuum image.

The \hi\ surface density shows a ``depression'' in galaxy A's position compared to the other systems, while the observed density enhancement surrounding the galaxy highlights the atomic gas counterpart of the ring. Galaxy B is detected within the velocity range \SIrange{675}{777}{\kms} (Figure \ref{fig:summary}c).
It seems to have retained its gas after the collision ($\mhi =  
\SI{4.04 \pm 0.07\e8}{\msun}$), though the peak of its \hi\ column density is offset from its optical counterpart, indicating a perturbed distribution likely caused by interaction with galaxy A. Moreover, a faint \hi\ plume, extending approximately four arc-minutes eastward from galaxy B and visible within the \SIrange{777}{900}{\kms} range (Figure \ref{fig:summary}b and \ref{fig:chmap}), further supports this interaction scenario.

Within the \SIrange{850}{900}{\kms} velocity range (Figures \ref{fig:summary}d and \ref{fig:chmap}), we detect the \hi\ counterpart of component C. Its atomic hydrogen distribution is also disturbed and seems to be spatially and kinematically connected with component D. 
A gas bridge connecting the two systems is detected in the velocity range \SIrange{887}{903}{\kms} shown in Figure \ref{fig:chmap}. At higher velocities, \SIrange{900}{1050}{\kms} (Figure \ref{fig:summary}e), it is possible to isolate component D. The densest part of its emission is marginally resolved in \hi\ and exhibits a extended and perturbed distribution towards the southwest surrounding this dense core.

WALLABY observations allow us to identify for the first time the neutral gas counterpart of the star-forming ring (Figure \ref{fig:summary}a). It is most clearly visible in the velocity range \SIrange{777}{900}{\kms} around the optical position of galaxy A (Figure \ref{fig:summary}d), and between \SIrange{805}{887}{\kms} (Figure \ref{fig:chmap}), where two high-density peaks can be identified moving along the ellipse that traces the optical position of the ring across different velocity channels. 

\subsection{H\textsc{i} kinematics}
Figure \ref{fig:vels} shows the \hi\ velocity and velocity dispersion maps of ESO 179-013. The large envelope of neutral gas that surrounds the system shows a disk-like velocity gradient with a pattern that is typical of rotation (Figure \ref{fig:vels}a). The only major break from the gradient is the peak in velocity (and also in velocity dispersion, Figure \ref{fig:vels}b) corresponding to component D. From SoFiA 2 we get the kinematic axis and show the position-velocity (PV) diagram obtained along it in Figure \ref{fig:vels}c. 

The global \hi\ velocity field also exhibits a clear warp at larger radii. To obtain a zeroth-order characterization of the geometric parameters of the kinematic twisting, we used the tilted-ring  code 3DBarolo \citep{diteodoro3DBAROLONew2015}, fixing the kinematic center to the optical position of Galaxy A and allowing the position angle to vary with radius. Within this simplified framework, we estimate a position angle variation of $\sim40^\circ$ across the \hi\ disk, indicative of significant kinematic twisting. Few other examples of collisional ring galaxies with evidence of warped outer disks are also known, such as UGC 7069 \citep{ghoshUGC7069Largest2008}, AM1354-250 \citep{connMUSEingsAM1354250Collisions2016}, and NGC~922 \citep{elagaliStudyCollisionalRing2018}.

The WALLABY data do not allow us to explore the velocity field of galaxy A's disk, as it is only marginally detected compared to the other structures. In the Cartwheel galaxy, 85\% of the detected \hi\ is concentrated in the outer ring, while a \hi\ ``hole'' is found at the location of the galaxy's nucleus \citep{higdonWheelsFireII1996}. The analogy with this system might explain the lack of a clear \hi\ detection in galaxy A.

According to the velocity distribution of H$\alpha$ knots \citep{parkerKathrynsWheelSpectacular2015} we expect some of the 21-cm emission up to $\sim\SI{900}{\kms}$ to be associated with the ring. To analyze this, we use the radio continuum image (Figure \ref{fig:radio}b) as a mask to select the star-forming knots in the H$\alpha$ image, avoiding confusion with galaxy C and component D in the southeast. We then fit an ellipse to the positions of these knots in the northeast part of the ring. Our best-fit ellipse to the H$\alpha$ data has $\sim100^{\prime\prime} \times 84^{\prime\prime}$ semi-axes and is overlaid in the panels of Figure \ref{fig:vels}. This ellipse is smaller than the one found by \citet{parkerKathrynsWheelSpectacular2015} ($127^{\prime\prime} \times 91^{\prime\prime}$), as they assume the H$\alpha$ ring extends as far as component D.
The physical diameter of the ring is of $\sim 4.8$ kpc. According to \cite{fiacconiAdaptiveMeshRefinement2012}, typical collisional rings have diameters between 6--10 kpc considering a Milky Way type of target hit by a gas deficient dwarf. The authors mention that the main factor on the size of the ring is the mass of the bullet and thus, the small dimension of the ring can be simply due to the small mass scale of the interacting galaxies.

The PV diagram extracted along this ellipse (Figure \ref{fig:vels}d) shows a sinusoidal velocity distribution similar to that traced by the H$\alpha$ knots in \citet{parkerKathrynsWheelSpectacular2015}. We measure an amplitude of $\sim\SI{58}{\kms}$, in excellent agreement with the \SI{60}{\kms} amplitude from optical data. However, due to our smaller inferred inclination from the optical data (33°), we derive a de-projected rotational velocity of \SI{70}{\kms}. Assuming rotation is the dominant velocity component, we estimate a dynamical mass of $\SI{2.7\e9}{\msun}$ within the ellipse, providing an upper limit on the total mass of galaxy A inside the ring.

In the PV diagram on the right, galaxy D appears as a vertical structure extending from 900 to $\sim\SI{1050}{\kms}$ at the location of component D around 200$^\circ$, further confirming its spatial and spectral association with the system. It is also visible in the PV diagram on the left, however, fainter since the kinematic axis does not pass through the densest part of the \hi\ distribution. This velocity pattern suggests the presence of non-rotational motions in galaxy D, indicating highly disturbed gas that is kinematically disconnected from the southwestern \hi\ envelope of the entire system.
Such features in PV diagrams are typically associated with outflows driven by tidal interactions or stellar feedback \citep{sancisiColdGasAccretion2008,kurapatiUncoveringExtraplanarGas2025,santanaMeerKATDiscoveryInfalling2025}. The velocity dispersion ($\sigma$) map (Figure \ref{fig:summary}b) shows values exceeding \SI{40}{\kms} in the same region, which is significantly higher than $\sim\SI{10}{\kms}$, typically observed in disks of dwarf galaxies \citep[e.g.,][]{iorioLITTLETHINGS3D2017,kurapatiStudiesExtremelyMetalpoor2024,mancerapinaGalaxyHaloConnection2025}. Such high values are consistent with turbulence, possibly resulting from interactions involving galaxy D and either galaxy C or A.

	\begin{table*}
\begin{center}
		\begin{tabular}{ccccllcc}
			\toprule
			\toprule
			& R.A. & Dec. &  $M_\star^b$ & $f_{\SI{1.4}{\giga\Hz}}$ & $L_{\SI{1.4}{\giga\Hz}}$ & SFR & SFR (H$\alpha$)$^a$  \\
			& hh:mm:ss & dd:mm:ss & ($\times10^8 M_\odot$) & ($\si{\milli\jansky\per\beam}$) &($\times10^{27}\si{\watt}$) & ($\si{\msun\per\yr}$) & ($\si{\msun\per\yr}$)\\
			\midrule
			A & 16:47:19.7$^a$ & $-$57:26:31$^a$ & $13.5\pm 0.2^c$ &$11.8\pm0.9$&$15.9\pm1.2$&$0.02\pm0.01$ & 0.01\\
			B & 16:47:27.8$^a$ & $-$57:25:35$^a$ & $2.7\pm 0.6$&$83.8\pm1.8$&$112.9\pm2.4$& $0.12\pm0.04$ & 0.08\\ 
			C & 16:47:15.5$^a$ & $-$57:26:44$^a$ & $1.2\pm0.3$&$65.2\pm1.8$&$87.9\pm2.4$& $0.09\pm0.03$ &\\
			D & 16:47:17.4     & $-$57:27:35    & $5.4\pm1.2$&$115.9\pm2.7$&$156.3\pm3.7$&$0.16\pm0.05$ &\\
			\bottomrule
		\end{tabular}
		\caption{Properties of the galaxies of the system. $^a$From \citet{parkerKathrynsWheelSpectacular2015}. $^b$These uncertainties are from adopting an accuracy of 0.1 dex in $\log M_\star$ \citep{jarrettNewWidefieldInfrared2023}. $^c$We use the AllWISE \citep{allwise} magnitude for this galaxy.}
		\label{tab:table}
\end{center}
	\end{table*}

\subsection{Star formation rates from radio continuum}\label{ref:sfr}

We estimate the star formation rates (SFRs) of the ESO 179-013 components using the 1.4 GHz radio continuum emission measured from the high-resolution WALLABY data, adopting the SFR--luminosity relation from \citet{filhoGlobalCorrelationsRadio2019}. At 1.4 GHz, the emission is dominated by synchrotron radiation associated with supernova activity from massive stars. Given the dwarf nature of these galaxies, the radio continuum is therefore expected to primarily trace star formation.

Galaxy D exhibits the highest radio continuum flux and consequently the highest SFR in the system: $\SI{0.16}{\msun\per\yr}$. Galaxies B and C have lower SFRs of $\SI{0.088}{\msun\per\yr}$ and $\SI{0.095 \pm 0.01}{\msun\per\yr}$, respectively. Galaxy B has a previous estimate of $\SI{0.08}{\msun\per\yr}$ \citep{parkerKathrynsWheelSpectacular2015}, in good agreement with our result.

Radio continuum emission from the central part of galaxy A is detected at the $4\sigma$ level, near a bright star-forming knot visible in both radio and optical images, located at the southwestern edge of its optical disk. This star-forming region (labeled as K in Figure \ref{fig:radio}) is associated with a compact \hi\ structure appearing in the \SIrange{852}{871}{\kms} range of the channel map (Figure \ref{fig:chmap}). It is unclear whether this knot and the gas cloud belong to galaxy A. If we consider only the central region of galaxy A, the inferred SFR is $\SI{0.020\pm0.009}{\msun\per\yr}$, which is consistent with the value found using H$\alpha$ \citep{parkerKathrynsWheelSpectacular2015}. Including the nearby knot increases the SFR to $\SI{0.108 \pm 0.035}{\msun\per\yr}$.
\section{Discussion}

\subsection{The \hi\ view of the H$\alpha$ ring}

The WALLABY data allows us to identify, for the first time, the \hi\ counterpart of the H$\alpha$ ring of ESO 179-013. The ring is not symmetric and it is visible as high-density peaks that follow the optical position of the ring in Figure \ref{fig:chmap}. The east side of the ring is clearly visible in the range \SIrange{805}{836}{\kms} while the west side is visible in the range \SIrange{836}{871}{\kms}. At higher velocities the \hi\ emission from the ring blends and overlaps with that of galaxy C. At lower velocities (\SIrange{758}{789}{\kms}), the northern side of the ring connects to the perturbed \hi\ disk of the ``intruder'' (galaxy B).

The collision velocity and the orientation of the intruder galaxy's disk relative to the target determine the ring's size and shape. To accurately derive the collision's impact parameters through simulations --- or to construct a detailed kinematic model of the system--- higher angular resolution \hi\ data are required. Consequently, with our current dataset, we can only offer a broad, qualitative discussion of the collision's nature. The systemic velocity of galaxy B derived from the \hi\ data is approximately \SI{745}{\kms}. Galaxy A was not clearly detected at 21 cm but assuming that its systemic velocity matches the midpoint of the ring's velocity range ($\sim\SI{845}{\kms}$), which corresponds to the systemic velocity of the entire \hi\ system \citep{koribalski1000BrightestHIPASS2004a}, then the radial component of the intruder's relative speed is about \SI{100}{\kms}. Galaxy B is located at a projected distance of 80$^{\prime\prime}$ (3.9 kpc) from galaxy A. This sets a lower limit of $\sim$ 38 Myr on the time since the collision occurred, assuming no transverse velocity component and using the projected separation.

The radio continuum emission is asymmetric, as the \hi\ distribution, and it is mostly confined to the western sides of the ring. A few radio continuum sources are also detected on the northern side of the ring, toward galaxy B, but no significant emission is found on the eastern side of the structure, where the ring gas density is also lower. The channel range \SIrange{805}{852}{\kms} in Figure \ref{fig:chmap}, as well as the integrated emission in Figure \ref{fig:summary}d, illustrate the higher column density of the northern and western halves of the ring.

\subsection{Galaxy D}

We identify a peak \hi\ column density of $7.0\times 10^{21}\mathrm{cm^{-2}}$ located at $16^{\rm h}47^{\rm m}17\fs38$ $-57\degr27\arcmin35\arcsec41$. This peak is not co-spatial with any known source and we refer to it as component D (Figure \ref{fig:summary}a). It is located to the south-east of galaxy A, with a separation of $\sim 72^{\prime\prime}$ ($\sim 3.5$ kpc). This source is also visible as a tail in the PV diagram of the system in Figure \ref{fig:vels}d  and dominates the emission at $> \SI{900}{\kms}$.

Integrating the \hi\ emission over the range \SIrange{900}{1050}{\kms}, we derive a \hi\ mass of $4.87\pm 0.09 \times 10^{8} M_\odot$. Additionally, it has the highest estimate of SFR from the continuum data, supporting the interpretation of this source being a galaxy. We estimate the stellar mass of galaxy D using data from the Wide-field Infrared Survey Explorer \citep[WISE;][]{wrightWidefieldInfraredSurvey2010a} WISE 1 band (\SI{3.4}{\micro\meter}).

Optical or infrared detection of object D is severely limited due to the presence of the foreground binary star HD 150915 and  2MASS J16472030-5727280, with $G$ band magnitudes of 7.68 and 11.15 mag, respectively \citep{gaiacollaborationGaiaEarlyData2021}. To mitigate contamination of the bright stars and overcrowding, we perform photometry using SExtractor \citep{bertinSExtractorSoftwareSource1996} on a detection image constructed from WISE W3 (12 $\upmu$m) observations.
We create the detection image via unsharp masking technique, i.e., we convolve the WISE 3 image with a Gaussian kernel ($\sigma =3\times$W3 PSF) and subtract from the original image to enhance the contrast of faint sources. We further use PSFEx \citep{bertinPSFEx2011} to model the point spread function (PSF) and improve photometric accuracy. Following this procedure, we measure a WISE 1 band magnitude of $10.13 \pm 0.01$ mag for galaxy D. Applying the empirical calibration of \citet{jarrettNewWidefieldInfrared2023} between WISE 1 luminosity and stellar mass, we obtain $M_* = 5.4 \times 10^8 M_\odot$. For this luminosity range, this is roughly equivalent to a 0.35 mass-to-light ratio with an expected accuracy for $\log M_\star$ is 0.1--0.2 dex (30\%--50\%). This mass-to-light matches well typical values found at the same wavelength in other gas-rich dwarfs using dynamical constraints \citep[e.g.,][]{mancerapinaGalaxyHaloConnection2025,marascoPhotometricDynamicMass2025}.

We note that the WISE 1 position of the source -- $16^{\rm h}47^{\rm m}16\fs6$ $-57\degr27\arcmin29\arcsec21$ -- does not coincide with the \hi\ peak, with an offset of $\SI{8.9}{\arcsecond}$. This offset could simply be due to spatial resolution effects, since the WISE PSF is significantly smaller than the ASKAP beam, and the measured offset is comparable to the pixel size of the radio image (\SI{7.5}{\arcsecond}). Moreover, given the possible gravitational interaction between the galaxies, such offsets are not unexpected: galaxy B also exhibits an offset between its optical and radio emission.
Far-UV observations have been used to confirm the position of other galaxies in similar scenarios \citep{cannonDiscoveryGasRichCompanion2014}, however, no suitable far-UV data is available for this region of the sky. Nevertheless, the radio continuum data allow us to identify that the star-forming component of galaxy D is extended ($\sim 2$ kpc), which encompasses both the WISE position and the \hi\ peak. Therefore we caution that our estimate should be considered as a lower limit to the stellar  mass of D, due to the incomplete recovery of the full 3.4 $\upmu$m emission affected by the binary stars.

\subsection{Alternatives to the collisional ring scenario}

Even though we detect the HI emission associated with the H$\alpha$ ring, this does not immediately corroborate the collisional ring scenario proposed for this system about ten years ago \citep{parkerKathrynsWheelSpectacular2015}. Several characteristics of the emission can put this scenario into question and we address the main ones in this section.

The most prominently unusual feature is the smoothness of the velocity field, which may imply that the \hi\ envelope is the gaseous disk of galaxy A, hosting different star-forming knots throughout it. As a consistency test, we consider the mass--diameter relation of \citet{wangNewLessonsSize2016} to probe if the whole \hi\ disk can be associated to only one object. For $M_\mathrm{HI} \sim \SI{2.25\e9}{\msun}$, the corresponding \hi\ disk diameter is expected to be $\SI{27.5 \pm 1.9}{\kpc}$. The diameter of the major axis at the $\SI{1}{\msun\persquare\pc}$ level is $\sim\SI{382}{\arcsec}$, which corresponds to $\sim\SI{18}{\kpc}$---a deviation 3 times larger than the typical scatter of the relation (0.06 dex). Coupled with the unusually high velocity dispersion, we conclude that the \hi\ envelope is not likely to be associated with a single galaxy.

Moreover, the ring exhibits an incomplete circular/elliptical morphology in both H$\alpha$ and \hi. This is not necessarily unexpected for collisional rings: off-center collisions occurring at large collision angles can produce asymmetric or partial ``C-shaped'' ring morphologies \citep{wongNGC922New2006,fiacconiAdaptiveMeshRefinement2012}. However, such a peculiar shape also raises the possibility that the SF knots previously interpreted as an incomplete collisional ring are instead diffuse star-forming regions triggered by the tidal interaction between galaxies A and B.

Another unusual feature is the small velocity difference between the target and the bullet inferred from the WALLABY observations ($\sim$ 100 km s$^{-1}$). Models aiming to reproduce the interactions leading to the formation of CRG typically assume larger relative velocities between  the intruder and the target, with values ranging between 250 km s$^{-1}$ and 650 km s$^{-1}$ \citep[e.g.,][]{fiacconiAdaptiveMeshRefinement2012}. Further hydrodynamical modeling is necessary to test these possibilities and determine whether a tidal encounter between galaxy A and B can better reproduce both the ring-like structure and the observed velocity field than a high-speed encounter between the two systems.

\subsection{A rare compact group of dwarfs in a void}

Few isolated groups composed exclusively of dwarf galaxies are known. \cite{stierwaltDirectEvidenceHierarchical2017} estimate that less than 5\% of dwarf galaxies are observed to have close companions in the SDSS main galaxy sample, even after accounting for incompleteness, identifying only seven such systems (one quintet, two quartets, and four triplets) with extents of $\sim 15$–80 kpc at distances of \SIrange{60}{200}{\mpc}. Recently, \cite{paudelDiscoveryRareGroup2024} discovered a new group of five dwarfs at 36 Mpc with an extension of 154 kpc. In both works, the groups were identified through visual inspection. Additional examples include a triplet in the Lynx–Cancer void \citep{chengalurDiscoveryExtremelyGas2013} with a separation of 25 kpc and the third companion of DDO 68 confirmed by \citet{correntiDDO68CHST2025} at a separation of 43 kpc. The existence of such groups provides a direct window to investigate hierarchical assembly and the origin of isolated intermediate-mass galaxies.

However, such optically selected samples likely suffer from significant incompleteness, as they are biased against diffuse, low-surface-brightness, gas-rich dwarfs and are affected by fiber collisions that limit the detection of close pairs. Using an untargeted interferometric H I survey that is largely insensitive to these effects, \citet{siljegGasrichDwarfGalaxy2025} find companion fractions of $\sim$10--20\% for gas-rich dwarfs, which is roughly three times higher than optical spectroscopic estimates, suggesting that dwarf multiples may be substantially underrepresented in optical studies. Therefore, nearby and well-resolved systems are valuable for studying the dynamical evolution and eventual fate of these groups of dwarfs in detail.

Compared to the known groups, ESO 179-013 is significantly closer, at 10 Mpc, and has a much smaller optical extent of $\sim 9~\mathrm{kpc}$ ($187^{\prime\prime}$). Based on the size, velocity coherence, and shared \hi\ envelope, we interpret ESO 179-013 as a compact group. Assuming a stellar mass of $\SI{2.4\e9}{\msun}$ for the system \citep{parkerKathrynsWheelSpectacular2015}, we estimate that ESO 179-013 may coalesce and give origin to a galaxy of at least $\log(M_\star / M_\odot)\sim 9.4$. Cosmological simulations of compact groups of dwarfs in TNG50 predict that they typically take $\leq$ 1 Gyr and up to $\sim$ 3 Gyr to fully merge, producing normally star-forming galaxies of $<10^{10} M_\odot$ at by $z=0$ \citep{flores-freitasCompactGroupsDwarf2024}. These compact groups are expected to be more common at high-redshift and possibly represent an important step in the growth of intermediate-mass galaxies. This shows that this system is important also as a proxy/analogue of processes and other compact groups at high redshift.

We also note the similarity of this system with the void triplets VGS\_38 \citep{kreckelOnlyLonelyImaging2011} and VGS\_31 \citep{beyguInteractingGalaxySystem2013}.
In both cases, the three galaxies are linearly aligned and embedded in a common \hi\ envelope with long bridges (50 kpc and 120 kpc, respectively), with fairly smooth rotation pattern. Both works argue that this might be tracing filamentary structures within the voids which are expected to be detected in \hi. This opens the possibility that deeper observations reaching low column density might reveal if ESO 179-013 is also tracing some substructure or even direct accretion of cold gas from a filament within a void.

\section{Summary}\label{sec:summary}
In this work, we presented the \hi\ observations of the triple system of dwarf galaxies ESO 179-013 obtained with the ASKAP telescope as part of the WALLABY pilot survey. This system, also known as Kathryn's Wheel, is the closest known ($10 \mathrm{~Mpc}$) example of a collisional ring galaxy. It is now revealed to be a compact group of at least four interacting dwarf galaxies (A–D) in the Local Void.
 
For the first time, we identify the \hi\ counterpart of the star-forming ring first detected in H$\alpha$ \citep{parkerKathrynsWheelSpectacular2015} and also detect it in the radio continuum. We estimate the ring velocity from the 21 cm observations and we confirm that the ring seems to be rotating with a velocity of $\sim\SI{58}{\kms}$ in agreement with spectral measurements from H$\alpha$ star-forming knots. We also report the discovery of a new galaxy in the system (Galaxy D), coincident with an \hi\ column density peak ($N_\mathrm{HI} \sim 7\times10^{21}$ cm$^{-2}$) and bright radio continuum emission, but undetected in optical and mid-infrared due to foreground stellar contamination. It shows the highest SFR in the group and its \hi\ morphology and kinematics show strong perturbations, suggesting that this system is likely interacting with the other members of the group.

When compared with other known compact groups and void dwarf systems, ESO 179-013 stands out due to its proximity and compact spatial configuration ($\sim$9 kpc). Its future coalescence could result in an intermediate-mass galaxy, consistent with predictions from simulations for hierarchical formation of galaxies. ESO 179-013 thus offers a rare, nearby laboratory to study interaction-induced processes in low-mass galaxies, group assembly, small scale clustering, and collisional ring formation physics. Future high-resolution observations will enable detailed kinematic modeling to constrain the collision geometry, investigate gas accretion or filamentary substructure within the Local Void, and determine the impact parameters of the collision. Chemical and dynamical analyses will be crucial to fully reconstruct the formation history of the system and its implications for galaxy evolution in low-density environments.

\section*{Acknowledgments}

RGS acknowledges support from CAPES (Coordena\c{c}\~ao de Aperfei\c{c}oamento de Pessoal de N\'ivel Superior) – Finance Code 001. MG acknowledges support from FAPERJ grant E-26/211.370/2021. DRG acknowledges FAPERJ (E-26/211.370/2021; E-26/211.527/2023) and CNPq (403011/2022-1; 315307/2023-4) grants. EC acknowledges financial support from INAF-Mini Grant RF-2023-SHAPES. RI acknowledges financial support from the grant CEX2021-001131-S funded by MICIU/AEI/ 10.13039/501100011033 and from the grant PID2021-123930OB-C21 funded by MICIU/AEI/ 10.13039/501100011033 and by ERDF/EU. PEMP is funded by the Dutch Research Council (NWO) through the Veni grant VI.Veni.222.364. KS acknowledges support from the Natural Sciences and Engineering Research Council of Canada (NSERC).

This scientific work uses data obtained from Inyarrimanha Ilgari Bundara, the CSIRO Murchison Radio-astronomy Observatory. We acknowledge the Wajarri Yamaji People as the Traditional Owners and native title holders of the Observatory site. CSIRO's ASKAP radio telescope is part of the Australia Telescope National Facility (\url{https://ror.org/05qajvd42}). Operation of ASKAP is funded by the Australian Government with support from the National Collaborative Research Infrastructure Strategy. ASKAP uses the resources of the Pawsey Supercomputing Research Centre. Establishment of ASKAP, Inyarrimanha Ilgari Bundara, the CSIRO Murchison Radio-astronomy Observatory and the Pawsey Supercomputing Research Centre are initiatives of the Australian Government, with support from the Government of Western Australia and the Science and Industry Endowment Fund.

This work was facilitated by the Australian SKA Regional Centre (AusSRC), Australia's portion of the international SKA Regional Centre Network (SRCNet), funded by the Australian Government through the Department of Industry, Science, and Resources (DISR; grant SKARC000001). AusSRC is an equal collaboration between CSIRO – Australia's national science agency, Curtin University, the Pawsey Supercomputing Research Centre, and the University of Western Australia.

This paper includes archived data obtained through the CSIRO ASKAP Science Data Archive, CASDA (\url{http://data.csiro.au}).

This publication makes use of data products from the Wide-field Infrared Survey Explorer, which is a joint project of the University of California, Los Angeles, and the Jet Propulsion Laboratory/California Institute of Technology, and NEOWISE, which is a project of the Jet Propulsion Laboratory/California Institute of Technology. WISE and NEOWISE are funded by the National Aeronautics and Space Administration.
%%%%%%%%%%%%%%%%%%%%%%%%%%%%%%%%%%%%%%%%%%%%%%%%%%


\bibliography{Bibliography}{}

\begin{thebibliography}{}
\expandafter\ifx\csname natexlab\endcsname\relax\def\natexlab#1{#1}\fi
\providecommand{\url}[1]{\href{#1}{#1}}
\providecommand{\dodoi}[1]{doi:~\href{http://doi.org/#1}{\nolinkurl{#1}}}
\providecommand{\doeprint}[1]{\href{http://ascl.net/#1}{\nolinkurl{http://ascl.net/#1}}}
\providecommand{\doarXiv}[1]{\href{https://arxiv.org/abs/#1}{\nolinkurl{https://arxiv.org/abs/#1}}}

\bibitem[{{Bertin}(2011)}]{bertinPSFEx2011}
{Bertin}, E. 2011, in Astronomical Society of the Pacific Conference Series,
  Vol. 442, Astronomical Data Analysis Software and Systems XX, ed. I.~N.
  {Evans}, A.~{Accomazzi}, D.~J. {Mink}, \& A.~H. {Rots}, 435

\bibitem[{Bertin \& Arnouts(1996)}]{bertinSExtractorSoftwareSource1996}
Bertin, E., \& Arnouts, S. 1996, \aaps, 117, 393, \dodoi{10.1051/aas:1996164}

\bibitem[{Besla {et~al.}(2018)Besla, Patton, Stierwalt, {Rodriguez-Gomez},
  Patel, Kallivayalil, Johnson, Pearson, Privon, \&
  Putman}]{beslaFrequencyDwarfGalaxy2018}
Besla, G., Patton, D.~R., Stierwalt, S., {et~al.} 2018, \mnras, 480, 3376,
  \dodoi{10.1093/mnras/sty2041}

\bibitem[{Beygu {et~al.}(2013)Beygu, Kreckel, {van de Weygaert}, {van der
  Hulst}, \& {van Gorkom}}]{beyguInteractingGalaxySystem2013}
Beygu, B., Kreckel, K., {van de Weygaert}, R., {van der Hulst}, J.~M., \& {van
  Gorkom}, J.~H. 2013, \aj, 145, 120, \dodoi{10.1088/0004-6256/145/5/120}

\bibitem[{{Boselli} \&
  {Gavazzi}(2006)}]{boselliEnvironmentalEffectsLateType2006}
{Boselli}, A., \& {Gavazzi}, G. 2006, \pasp, 118, 517, \dodoi{10.1086/500691}

\bibitem[{Cannon {et~al.}(2014)Cannon, Johnson, McQuinn, Alfvin, Bailin, Ford,
  Girardi, Hirschauer, Janowiecki, Salzer, Van~Sistine, Dolphin, Elson,
  Koribalski, Marigo, Rosenberg, Rosenfield, Skillman, Venkatesan, \&
  Warren}]{cannonDiscoveryGasRichCompanion2014}
Cannon, J.~M., Johnson, M., McQuinn, K. B.~W., {et~al.} 2014, \apjl, 787, L1,
  \dodoi{10.1088/2041-8205/787/1/L1}

\bibitem[{Chengalur \& Pustilnik(2013)}]{chengalurDiscoveryExtremelyGas2013}
Chengalur, J.~N., \& Pustilnik, S.~A. 2013, \mnras, 428, 1579,
  \dodoi{10.1093/mnras/sts138}

\bibitem[{Conn {et~al.}(2016)Conn, Fogarty, Smith, \&
  Candlish}]{connMUSEingsAM1354250Collisions2016}
Conn, B.~C., Fogarty, L. M.~R., Smith, R., \& Candlish, G.~N. 2016, \apj, 819,
  165, \dodoi{10.3847/0004-637X/819/2/165}

\bibitem[{Correnti {et~al.}(2025)Correnti, Annibali, Bellazzini, Marinelli,
  Aloisi, Cignoni, Tosi, Pascale, Cannon, Schisgal, Hunt, Sacchi, \&
  Sohn}]{correntiDDO68CHST2025}
Correnti, M., Annibali, F., Bellazzini, M., {et~al.} 2025, \apj, 982, 31,
  \dodoi{10.3847/1538-4357/adb7e6}

\bibitem[{Courtois {et~al.}(2023)Courtois, Said, Mould, Jarrett, Pomar{\`e}de,
  Westmeier, {Staveley-Smith}, Dupuy, Hong, Guinet, Howlett, Deg, For, Kleiner,
  Koribalski, {Lee-Waddell}, Rhee, Spekkens, Wang, Wong, Bigiel, Bosma,
  Colless, Davis, Holwerda, Karachentsev, {Kraan-Korteweg}, McQuinn, Meurer,
  Obreschkow, \& Taylor}]{courtoisWALLABYPrepilotPilot2023}
Courtois, H.~M., Said, K., Mould, J., {et~al.} 2023, \mnras, 519, 4589,
  \dodoi{10.1093/mnras/stac3246}

\bibitem[{Di~Teodoro \& Fraternali(2015)}]{diteodoro3DBAROLONew2015}
Di~Teodoro, E.~M., \& Fraternali, F. 2015, \mnras, 451, 3021,
  \dodoi{10.1093/mnras/stv1213}

\bibitem[{Elagali {et~al.}(2018)Elagali, Wong, Oh, {Staveley-Smith},
  Koribalski, Bekki, \& Zwaan}]{elagaliStudyCollisionalRing2018}
Elagali, A., Wong, O.~I., Oh, S.-H., {et~al.} 2018, \mnras, 476, 5681,
  \dodoi{10.1093/mnras/sty741}

\bibitem[{Fiacconi {et~al.}(2012)Fiacconi, Mapelli, Ripamonti, \&
  Colpi}]{fiacconiAdaptiveMeshRefinement2012}
Fiacconi, D., Mapelli, M., Ripamonti, E., \& Colpi, M. 2012, \mnras, 425, 2255,
  \dodoi{10.1111/j.1365-2966.2012.21566.x}

\bibitem[{Filho {et~al.}(2019)Filho, Tabatabaei, S{\'a}nchez~Almeida,
  {Mu{\~n}oz-Tu{\~n}{\'o}n}, \& Elmegreen}]{filhoGlobalCorrelationsRadio2019}
Filho, M.~E., Tabatabaei, F.~S., S{\'a}nchez~Almeida, J.,
  {Mu{\~n}oz-Tu{\~n}{\'o}n}, C., \& Elmegreen, B.~G. 2019, Monthly Notices of
  the Royal Astronomical Society, 484, 543, \dodoi{10.1093/mnras/sty3199}

\bibitem[{{Flores-Freitas} {et~al.}(2024){Flores-Freitas}, Trevisan,
  M{\"u}ckler, Mamon, {Schnorr-M{\"u}ller}, \&
  Bootz}]{flores-freitasCompactGroupsDwarf2024}
{Flores-Freitas}, R., Trevisan, M., M{\"u}ckler, M., {et~al.} 2024, \mnras,
  528, 5804, \dodoi{10.1093/mnras/stae367}

\bibitem[{{Gaia Collaboration} {et~al.}(2021){Gaia Collaboration}, Brown,
  Vallenari, Prusti, {de Bruijne}, Babusiaux, Biermann, Creevey, Evans, Eyer,
  Hutton, Jansen, Jordi, Klioner, Lammers, Lindegren, Luri, Mignard, Panem,
  Pourbaix, Randich, Sartoretti, Soubiran, Walton, Arenou, {Bailer-Jones},
  Bastian, Cropper, Drimmel, Katz, Lattanzi, {van Leeuwen}, Bakker, Cacciari,
  Casta{\~n}eda, De~Angeli, Ducourant, Fabricius, Fouesneau, Fr{\'e}mat,
  Guerra, Guerrier, Guiraud, {Jean-Antoine Piccolo}, Masana, Messineo, Mowlavi,
  Nicolas, Nienartowicz, Pailler, Panuzzo, Riclet, Roux, Seabroke, Sordo,
  Tanga, Th{\'e}venin, {Gracia-Abril}, Portell, Teyssier, Altmann, Andrae,
  {Bellas-Velidis}, Benson, Berthier, Blomme, Brugaletta, Burgess, Busso,
  Carry, Cellino, Cheek, Clementini, Damerdji, Davidson, Delchambre, Dell'Oro,
  {Fern{\'a}ndez-Hern{\'a}ndez}, Galluccio, {Garc{\'i}a-Lario},
  {Garcia-Reinaldos}, {Gonz{\'a}lez-N{\'u}{\~n}ez}, Gosset, Haigron, Halbwachs,
  Hambly, Harrison, Hatzidimitriou, Heiter, Hern{\'a}ndez, Hestroffer, Hodgkin,
  Holl, Jan{\ss}en, {Jevardat de Fombelle}, Jordan, {Krone-Martins}, Lanzafame,
  L{\"o}ffler, Lorca, Manteiga, Marchal, Marrese, Moitinho, Mora, Muinonen,
  Osborne, Pancino, Pauwels, Petit, {Recio-Blanco}, Richards, Riello,
  Rimoldini, Robin, Roegiers, Rybizki, Sarro, Siopis, Smith, Sozzetti, Ulla,
  Utrilla, {van Leeuwen}, {van Reeven}, Abbas, Abreu~Aramburu, Accart, Aerts,
  Aguado, Ajaj, Altavilla, {\'A}lvarez, {\'A}lvarez Cid-Fuentes, Alves,
  Anderson, Anglada~Varela, Antoja, Audard, Baines, Baker,
  {Balaguer-N{\'u}{\~n}ez}, Balbinot, Balog, Barache, Barbato, Barros, Barstow,
  Bartolom{\'e}, Bassilana, Bauchet, {Baudesson-Stella}, Becciani, Bellazzini,
  Bernet, Bertone, Bianchi, {Blanco-Cuaresma}, Boch, Bombrun, Bossini,
  Bouquillon, Bragaglia, Bramante, Breedt, Bressan, Brouillet, Bucciarelli,
  Burlacu, Busonero, Butkevich, Buzzi, Caffau, Cancelliere, C{\'a}novas,
  {Cantat-Gaudin}, Carballo, Carlucci, Carnerero, Carrasco, Casamiquela,
  Castellani, {Castro-Ginard}, Castro~Sampol, Chaoul, Charlot, Chemin,
  Chiavassa, Cioni, Comoretto, Cooper, Cornez, Cowell, Crifo, Crosta, Crowley,
  Dafonte, Dapergolas, David, \& David}]{gaiacollaborationGaiaEarlyData2021}
{Gaia Collaboration}, Brown, A. G.~A., Vallenari, A., {et~al.} 2021, \aap, 649,
  A1, \dodoi{10.1051/0004-6361/202039657}

\bibitem[{Ghosh \& Mapelli(2008)}]{ghoshUGC7069Largest2008}
Ghosh, K.~K., \& Mapelli, M. 2008, \mnras, 386, L38,
  \dodoi{10.1111/j.1745-3933.2008.00456.x}

\bibitem[{Higdon(1996)}]{higdonWheelsFireII1996}
Higdon, J.~L. 1996, \apj, 467, 241, \dodoi{10.1086/177599}

\bibitem[{Hotan {et~al.}(2021)Hotan, Bunton, Chippendale, Whiting, Tuthill,
  Moss, McConnell, Amy, Huynh, Allison, Anderson, Bannister, Bastholm,
  Beresford, Bock, Bolton, Chapman, Chow, Collier, Cooray, Cornwell, Diamond,
  Edwards, Feain, Franzen, George, Gupta, Hampson, {Harvey-Smith}, Hayman,
  Heywood, Jacka, Jackson, Jackson, Jeganathan, Johnston, Kesteven, Kleiner,
  Koribalski, {Lee-Waddell}, Lenc, Lensson, Mackay, Mahony,
  {McClure-Griffiths}, McConigley, Mirtschin, Ng, Norris, Pearce, Phillips,
  Pilawa, Raja, Reynolds, Roberts, Roxby, Sadler, Shields, Schinckel, Serra,
  Shaw, Sweetnam, Troup, Tzioumis, Voronkov, \&
  Westmeier}]{hotanAustralianSquareKilometre2021}
Hotan, A.~W., Bunton, J.~D., Chippendale, A.~P., {et~al.} 2021, \pasa, 38,
  e009, \dodoi{10.1017/pasa.2021.1}

\bibitem[{{Iorio} {et~al.}(2017){Iorio}, {Fraternali}, {Nipoti}, {Di Teodoro},
  {Read}, \& {Battaglia}}]{iorioLITTLETHINGS3D2017}
{Iorio}, G., {Fraternali}, F., {Nipoti}, C., {et~al.} 2017, \mnras, 466, 4159,
  \dodoi{10.1093/mnras/stw3285}

\bibitem[{Jarrett {et~al.}(2023)Jarrett, Cluver, Taylor, Bellstedt, Robotham,
  \& Yao}]{jarrettNewWidefieldInfrared2023}
Jarrett, T.~H., Cluver, M.~E., Taylor, E.~N., {et~al.} 2023, \apj, 946, 95,
  \dodoi{10.3847/1538-4357/acb68f}

\bibitem[{Koribalski {et~al.}(2004)Koribalski, {Staveley-Smith}, Kilborn,
  Ryder, {Kraan-Korteweg}, {Ryan-Weber}, Ekers, Jerjen, Henning, Putman, Zwaan,
  {de Blok}, Calabretta, Disney, Minchin, Bhathal, Boyce, Drinkwater, Freeman,
  Gibson, Green, Haynes, Juraszek, Kesteven, Knezek, Mader, Marquarding, Meyer,
  Mould, Oosterloo, O'Brien, Price, Sadler, Schr{\"o}der, Stewart, Stootman,
  Waugh, Warren, Webster, \& Wright}]{koribalski1000BrightestHIPASS2004a}
Koribalski, B.~S., {Staveley-Smith}, L., Kilborn, V.~A., {et~al.} 2004, \aj,
  128, 16, \dodoi{10.1086/421744}

\bibitem[{Koribalski {et~al.}(2020)Koribalski, {Staveley-Smith}, Westmeier,
  Serra, Spekkens, Wong, {Lee-Waddell}, Lagos, Obreschkow, {Ryan-Weber}, Zwaan,
  Kilborn, Bekiaris, Bekki, Bigiel, Boselli, Bosma, Catinella, Chauhan, Cluver,
  Colless, Courtois, Crain, {de Blok}, D{\'e}nes, Duffy, Elagali, Fluke, For,
  Heald, Henning, Hess, Holwerda, Howlett, Jarrett, Jones, Jones, J{\'o}zsa,
  Jurek, J{\"u}tte, Kamphuis, Karachentsev, Kerp, Kleiner, {Kraan-Korteweg},
  {L{\'o}pez-S{\'a}nchez}, Madrid, Meyer, Mould, Murugeshan, Norris, Oh,
  Oosterloo, Popping, Putman, Reynolds, Rhee, Robotham, Ryder, Schr{\"o}der,
  Shao, Stevens, Taylor, {van{\^A} der Hulst}, {Verdes-Montenegro}, Wakker,
  Wang, Whiting, Winkel, \& Wolf}]{koribalskiWALLABYSKAPathfinder2020}
Koribalski, B.~S., {Staveley-Smith}, L., Westmeier, T., {et~al.} 2020, \apss,
  365, 118, \dodoi{10.1007/s10509-020-03831-4}

\bibitem[{Kourkchi {et~al.}(2020)Kourkchi, Courtois, Graziani, Hoffman,
  Pomar{\`e}de, Shaya, \& Tully}]{kourkchiCosmicflows3TwoDistance2020}
Kourkchi, E., Courtois, H.~M., Graziani, R., {et~al.} 2020, \aj, 159, 67,
  \dodoi{10.3847/1538-3881/ab620e}

\bibitem[{Kreckel {et~al.}(2011)Kreckel, Platen, {Arag{\'o}n-Calvo}, {van
  Gorkom}, {van de Weygaert}, {van der Hulst}, Kova{\v c}, Yip, \&
  Peebles}]{kreckelOnlyLonelyImaging2011}
Kreckel, K., Platen, E., {Arag{\'o}n-Calvo}, M.~A., {et~al.} 2011, \aj, 141, 4,
  \dodoi{10.1088/0004-6256/141/1/4}

\bibitem[{Kurapati {et~al.}(2024)Kurapati, Pustilnik, \&
  Egorova}]{kurapatiStudiesExtremelyMetalpoor2024}
Kurapati, S., Pustilnik, S.~A., \& Egorova, E.~S. 2024, \mnras, 533, 1178,
  \dodoi{10.1093/mnras/stae1894}

\bibitem[{Kurapati {et~al.}(2025)Kurapati, Pisano, {de~Blok}, Kamphuis, Zabel,
  {de~Villiers}, Healy, Maccagni, Kleiner, Adams, Amram, Athanassoula, Bigiel,
  Bosma, Brinks, Chemin, Combes, Dettmar, J{\'o}zsa, Koribalski, Marasco,
  Meurer, Mogotsi, Mohapatra, Rajohnson, Schinnerer, Sorgho, Spekkens,
  {Verdes-Montenegro}, Veronese, \&
  Walter}]{kurapatiUncoveringExtraplanarGas2025}
Kurapati, S., Pisano, D.~J., {de~Blok}, W. J.~G., {et~al.} 2025, \mnras, 538,
  1272, \dodoi{10.1093/mnras/staf387}

\bibitem[{Lynds \& Toomre(1976)}]{lyndsInterpretationRingGalaxies1976}
Lynds, R., \& Toomre, A. 1976, \apj, 209, 382, \dodoi{10.1086/154730}

\bibitem[{Madore {et~al.}(2009)Madore, Nelson, \&
  Petrillo}]{madoreAtlasCatalogCollisional2009}
Madore, B.~F., Nelson, E., \& Petrillo, K. 2009, \apjs, 181, 572,
  \dodoi{10.1088/0067-0049/181/2/572}

\bibitem[{{Mancera Pi{\~n}a} {et~al.}(2025){Mancera Pi{\~n}a}, {Read}, {Kim},
  {Marasco}, {Benavides}, {Glowacki}, {Pezzulli}, \&
  {Lagos}}]{mancerapinaGalaxyHaloConnection2025}
{Mancera Pi{\~n}a}, P.~E., {Read}, J.~I., {Kim}, S., {et~al.} 2025, \aap, 699,
  A311, \dodoi{10.1051/0004-6361/202554381}

\bibitem[{Mapelli {et~al.}(2008)Mapelli, Moore, Ripamonti, Mayer, Colpi, \&
  Giordano}]{mapelliAreRingGalaxies2008}
Mapelli, M., Moore, B., Ripamonti, E., {et~al.} 2008, \mnras, 383, 1223,
  \dodoi{10.1111/j.1365-2966.2007.12650.x}

\bibitem[{{Marasco} {et~al.}(2025){Marasco}, {Fall}, {Di Teodoro}, \& {Mancera
  Pi{\~n}a}}]{marascoPhotometricDynamicMass2025}
{Marasco}, A., {Fall}, S.~M., {Di Teodoro}, E.~M., \& {Mancera Pi{\~n}a}, P.~E.
  2025, \aap, 695, L23, \dodoi{10.1051/0004-6361/202553925}

\bibitem[{Murugeshan {et~al.}(2023)Murugeshan, D{\v z}ud{\v z}ar, Bagge,
  O'Beirne, Wong, Kilborn, Cluver, Lutz, \&
  Elagali}]{murugeshanHiRingGalaxies2023}
Murugeshan, C., D{\v z}ud{\v z}ar, R., Bagge, R., {et~al.} 2023, \pasa, 40,
  e018, \dodoi{10.1017/pasa.2023.19}

\bibitem[{O'Beirne {et~al.}(2024)O'Beirne, {Staveley-Smith}, Wong, Westmeier,
  Batten, Kilborn, {Lee-Waddell}, Mancera~Pi{\~n}a, Rom{\'a}n,
  {Verdes-Montenegro}, Catinella, Cortese, Deg, D{\'e}nes, For, Kamphuis,
  Koribalski, Murugeshan, Rhee, Spekkens, Wang, Bekki, \&
  {L{\'p}pez-S{\'a}nchez}}]{obeirneWALLABYPilotSurvey2024}
O'Beirne, T., {Staveley-Smith}, L., Wong, O.~I., {et~al.} 2024, \mnras, 528,
  4010, \dodoi{10.1093/mnras/stae215}

\bibitem[{Paliya \& Saikia(2024)}]{paliyaGRayemittingCollisionalRing2024}
Paliya, V.~S., \& Saikia, D.~J. 2024, \apjl, 967, L26,
  \dodoi{10.3847/2041-8213/ad4999}

\bibitem[{Parker {et~al.}(2015)Parker, Zijlstra, Stupar, Cluver, Frew, Bendo,
  \& Boji{\v c}i{\'c}}]{parkerKathrynsWheelSpectacular2015}
Parker, Q.~A., Zijlstra, A.~A., Stupar, M., {et~al.} 2015, \mnras, 452, 3759,
  \dodoi{10.1093/mnras/stv1432}

\bibitem[{Paudel {et~al.}(2024)Paudel, Sabiu, Yoon, Duc, Yoo, \&
  M{\"u}ller}]{paudelDiscoveryRareGroup2024}
Paudel, S., Sabiu, C.~G., Yoon, S.-J., {et~al.} 2024, \apjl, 976, L18,
  \dodoi{10.3847/2041-8213/ad8f3c}

\bibitem[{{Renaud} {et~al.}(2018){Renaud}, {Athanassoula}, {Amram}, {Bosma},
  {Bournaud}, {Duc}, {Epinat}, {Fensch}, {Kraljic}, {Perret}, \&
  {Struck}}]{renaudMorphologyEnhancedStar2018}
{Renaud}, F., {Athanassoula}, E., {Amram}, P., {et~al.} 2018, \mnras, 473, 585,
  \dodoi{10.1093/mnras/stx2360}

\bibitem[{Sancisi {et~al.}(2008)Sancisi, Fraternali, Oosterloo, \& {van der
  Hulst}}]{sancisiColdGasAccretion2008}
Sancisi, R., Fraternali, F., Oosterloo, T., \& {van der Hulst}, T. 2008, \aapr,
  15, 189, \dodoi{10.1007/s00159-008-0010-0}

\bibitem[{Santana {et~al.}(2025)Santana, Maccagni, Deane, \&
  Healy}]{santanaMeerKATDiscoveryInfalling2025}
Santana, K.~C., Maccagni, F.~M., Deane, R.~P., \& Healy, J. 2025, \mnras, 540,
  2396, \dodoi{10.1093/mnras/staf832}

\bibitem[{{\v S}iljeg {et~al.}(2025){\v S}iljeg, Adams, Fraternali, Hess,
  Marasco, D{\'e}nes, Garrido, Lucero, Morganti, {S{\'a}nchez-Exp{\'o}sito}, \&
  {van der Hulst}}]{siljegGasrichDwarfGalaxy2025}
{\v S}iljeg, B., Adams, E. A.~K., Fraternali, F., {et~al.} 2025, Astronomy and
  Astrophysics, 703, A295, \dodoi{10.1051/0004-6361/202554841}

\bibitem[{Stierwalt {et~al.}(2017)Stierwalt, Liss, Johnson, Patton, Privon,
  Besla, Kallivayalil, \& Putman}]{stierwaltDirectEvidenceHierarchical2017}
Stierwalt, S., Liss, S.~E., Johnson, K.~E., {et~al.} 2017, Nature Astronomy, 1,
  1, \dodoi{10.1038/s41550-016-0025}

\bibitem[{Tully {et~al.}(2016)Tully, Courtois, \&
  Sorce}]{tullyCosmicflows32016}
Tully, R.~B., Courtois, H.~M., \& Sorce, J.~G. 2016, \aj, 152, 50,
  \dodoi{10.3847/0004-6256/152/2/50}

\bibitem[{Tully {et~al.}(2019)Tully, Pomar{\`e}de, Graziani, Courtois, Hoffman,
  \& Shaya}]{tullyCosmicflows3CosmographyLocal2019}
Tully, R.~B., Pomar{\`e}de, D., Graziani, R., {et~al.} 2019, \apj, 880, 24,
  \dodoi{10.3847/1538-4357/ab2597}

\bibitem[{Wang {et~al.}(2016)Wang, Koribalski, Serra, {van der Hulst},
  Roychowdhury, Kamphuis, \& N.~Chengalur}]{wangNewLessonsSize2016}
Wang, J., Koribalski, B.~S., Serra, P., {et~al.} 2016, \mnras, 460, 2143,
  \dodoi{10.1093/mnras/stw1099}

\bibitem[{Westmeier {et~al.}(2021)Westmeier, Kitaeff, Pallot, Serra, {van der
  Hulst}, Jurek, Elagali, For, Kleiner, Koribalski, {Lee-Waddell}, Mould,
  Reynolds, Rhee, \& {Staveley-Smith}}]{westmeierSOFIA2Automated2021}
Westmeier, T., Kitaeff, S., Pallot, D., {et~al.} 2021, \mnras, 506, 3962,
  \dodoi{10.1093/mnras/stab1881}

\bibitem[{Westmeier {et~al.}(2022)Westmeier, Deg, Spekkens, Reynolds, Shen,
  Gaudet, Goliath, Huynh, Venkataraman, Lin, O'Beirne, Catinella, Cortese,
  D{\'e}nes, Elagali, For, J{\'o}zsa, Howlett, {van der Hulst}, Jurek,
  Kamphuis, Kilborn, Kleiner, Koribalski, {Lee-Waddell}, Murugeshan, Rhee,
  Serra, Shao, {Staveley-Smith}, Wang, Wong, Zwaan, Allison, Anderson, Ball,
  Bock, Brodrick, Bunton, Cooray, Gupta, Hayman, Mahony, Moss, Ng, Pearce,
  Raja, Roxby, Voronkov, Warhurst, Courtois, \&
  Said}]{westmeierWALLABYPilotSurvey2022}
Westmeier, T., Deg, N., Spekkens, K., {et~al.} 2022, \pasa, 39, e058,
  \dodoi{10.1017/pasa.2022.50}

\bibitem[{Wong {et~al.}(2006)Wong, Meurer, Bekki, Hanish, Kennicutt,
  {Bland-Hawthorn}, {Ryan-Weber}, Koribalski, Kilborn, Putman, Heiner, Webster,
  Allen, Dopita, Doyle, Drinkwater, Ferguson, Freeman, Heckman, Hoopes, Knezek,
  Meyer, Oey, Seibert, Smith, {Staveley-Smith}, Thilker, Werk, \&
  Zwaan}]{wongNGC922New2006}
Wong, O.~I., Meurer, G.~R., Bekki, K., {et~al.} 2006, \mnras, 370, 1607,
  \dodoi{10.1111/j.1365-2966.2006.10589.x}

\bibitem[{Wright {et~al.}(2010)Wright, Eisenhardt, Mainzer, Ressler, Cutri,
  Jarrett, Kirkpatrick, Padgett, McMillan, Skrutskie, Stanford, Cohen, Walker,
  Mather, Leisawitz, Gautier, McLean, Benford, Lonsdale, Blain, Mendez, Irace,
  Duval, Liu, Royer, Heinrichsen, Howard, Shannon, Kendall, Walsh, Larsen,
  Cardon, Schick, Schwalm, Abid, Fabinsky, Naes, \&
  Tsai}]{wrightWidefieldInfraredSurvey2010a}
Wright, E.~L., Eisenhardt, P. R.~M., Mainzer, A.~K., {et~al.} 2010, \aj, 140,
  1868, \dodoi{10.1088/0004-6256/140/6/1868}

\bibitem[{Wright {et~al.}(2019)Wright, Eisenhardt, Mainzer, Ressler, Cutri,
  Jarrett, Kirkpatrick, Padgett, McMillan, Skrutskie, Stanford, Cohen, Walker,
  Mather, Leisawitz, Gautier, McLean, Benford, Lonsdale, Blain, Mendez, Irace,
  Duval, Liu, Royer, Heinrichsen, Howard, Shannon, Kendall, Walsh, Larsen,
  Cardon, Schick, Schwalm, Abid, Fabinsky, Naes, \& Tsai}]{allwise}
---. 2019, AllWISE Source Catalog,  IPAC, \dodoi{10.26131/IRSA1}

\end{thebibliography}
\bibliographystyle{aasjournal}



%%%%%%%%%%%%%%%%%%%%%%%%%%%%%%%%%%%%%%%%%%%%%%%%%%

%%%%%%%%%%%%%%%%% APPENDICES %%%%%%%%%%%%%%%%%%%%%

\end{document}